\documentclass[aps,prd,twocolumn,showpacs,superscriptaddress,groupedaddress]{revtex4}

\usepackage{graphicx}
\usepackage{amsmath,amssymb}
\usepackage{hyperref}

\begin{document}

\title{Deriving Einstein's Field Equation(EFE) and Modified Gravity by Statistical Mechanics and Quantization of Non-Commuting Spaces}
\author{R. Hassannejad}
\email{r.hassannejad@ut.ac.ir}
\affiliation{Department of Physics, University of Tehran, Tehran 14395-547, Iran}
\author{S. Navid Mousavi}
\email{s.n.mousavi@shirazu.ac.ir}
\affiliation{Department of Physics, Shiraz University, Shiraz 71454, Iran.}
\date{\today}

\begin{abstract}
In this paper, we derive the Einstein's field equation (EFE) by considering an non-commuting two dimensional quantized space, which can be excited by absorbing energy. Any variation of the energy level of space quantas, will result in a change in the quanta's area state. This means, that the geometry of space depends on the energy which exists in it. Using Holographic principle and Unruh effect, without any further assumptions, our model leads to Newton's law of gravity in the limit $\hbar \to 0$.
\end{abstract}
\maketitle

\section{Introduction}

After the introduction of general relativity by Einstein, and its experimental affirmations, together with the development of quantum mechanics and its great achievements, many efforts were made to connect these two theories to answer questions about the nature of gravity. In this process, physicists made great theories such as string theory \cite{blum2015brief, damour1994string, seiberg1999string}, and loop quantum gravity theory \cite{rovelli2008loop,carlip2001quantum}.

There has always been a quest to find a connection between thermodynamics and gravity, which was first done by Beckenstein in 1973 \cite{1}, and Hawking in 1974\cite{2}. They compared the mass, electric charge, and angular momentum of black-holes with thermodynamical parameters, and found relations for temperature $T$, and entropy $S$, of black-holes. Also, theorists tried to drive the Einstein's field equation using thermodynamics, and we can point out the work done by Jacobson in 1995 \cite{9}, where he uses the first law of thermodynamics to derive the Einstein's field equation. In a brilliant paper in 2011 \cite{11}, Verlinde used the Holographic principle and equipartition theorem to derive Newton's law of gravity and Einstein's equation. In fact, just as quantum mechanics was started by the study of black body radiation problem in a statistical mechanics framework, the Holographic principle came to existence as a consequence of statistical mechanics of black holes\cite{bousso2002holographic}.

Considering the things done in order to connect quantum mechanics and gravitation, quantization of space seems to be a good approach. There exist different methods to quantize the space, but in general, some variable which is commutative in the continuous limit assumed to be non-commuting for the sake of forming a different lie algebra \cite{gouba2016comparative}. This idea was first introduced by Heisenberg to solve the infinities in the quantum field theory\cite{gouba2016comparative}.
The first concrete example of a Lorentz discrete space-time was made by Snyder in 1947 \cite{snyder1947quantized}, and after that, in the 1980s Connes developed the noncommutative differential geometry \cite{connes1985non}. A more detailed history of the subject is written in the reference \cite{akofor2008quantum}.

Romero \textit{et al.} in a paper in 2003 \cite{7}, calculated the area of a two dimensional non-commuting space with non-commutation relation $[\hat{x}_1,\hat{x}_2] = i \hbar \Theta, \quad \Theta > 0$. They found the below result for the area of the space
\begin{equation}\label{eq-2}
A_n = 2 \pi \hbar \Theta (n + \frac{1}{2}),\quad n = 0, 1, 2, \cdots
\end{equation}
and compared their result with the area calculated in the reference \cite{bekenstein1974quantum} for the black hole horizon, which can be written as $A_n = \tilde{\gamma} l_p^2 n$, with $n = 1, 2, 3, \cdots$, and also the loop quantum gravity theory result for the space area \cite{alekseev2003area}, with the form $
A(j_i) = \gamma l_p^2 (j_i + \frac{1}{2})$. Considering the similarity between the results given by Romero \textit{et al.} \cite{7}, and Alekseev \textit{et al.} \cite{alekseev2003area}, which are driven with different approaches, we can conclude that non-commuting geometry and quantum gravity must be connected \cite{7}

The main idea behind this paper is the quantization of non-commuting space and the assumption of excitation in space quantas when some sort of energy is in the space. \textit{i.e.,} we consider the space containing $N$ quantas, where \emph{each quanta can be excited by absorbing energy}. This excitation changes the quanta's area state, and the ultimate space geometry should be determined by the sum over area states of the space quantas.

When there exists no energy or mass in the space, all the quantas will have the same area (Eq.\eqref{eq-2} with $n = 0$). This situation can be thought of similar to the ground state of a system of harmonic oscillators, which all of them have the same energy state $E = \frac{1}{2} \hbar \omega$ ($n = 0$). Just by inserting some energy into space, space quantas will become excited and the area of each quanta would change as the Eq.\eqref{eq-2} dictates. Each quanta will have a specific area, which is dependent on its energy. Fig.\eqref{fig1}, presents a schematic view of a flat space with all quantas in the ground area state ($n = 0$) with an area equal to $\hbar \Theta$, but in Fig.\eqref{fig2} you can see the space quantas which is excited and each quanta has a specific area.

With these descriptions in mind, we derive the EFE, modified gravity equation, and Newton's law of gravity. The procedure is presented in three parts. First in part \eqref{part1}, we calculate the entropy and the internal energy of the space, then in part \eqref{part2}, the derivation of the EFE and modified gravity will be presented, and finally, in part \eqref{part3}, we show that our model yields Newton's law of gravity using Holographic principle and Unruh effect.

\section{Procedure}
\subsection{Entropy and internal energy of space}\label{part1}
\begin{figure}
\centering
\includegraphics[width = 6cm, height= 6cm]{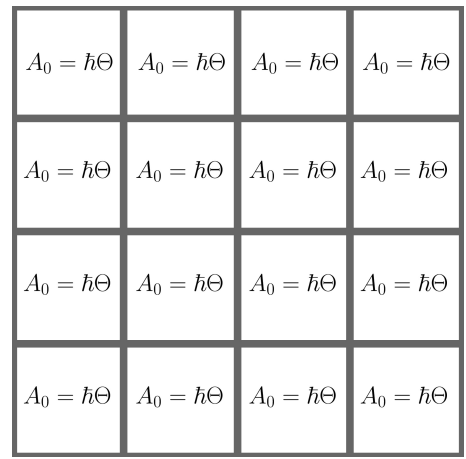}
\caption{Schematic view of the space quantas in the ground state. As it can be seen, each quanta has an area equal to $\hbar \Theta$, and the total area of $16\hbar \Theta$ is filled with $16$ quantas.}
\label{fig1}
\end{figure}

\begin{figure}
\centering
\includegraphics[width = 6cm, height= 6cm]{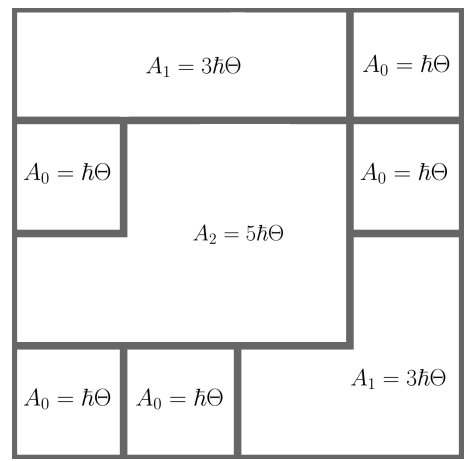}
\caption{Schematic view of the excited space quantas. As it can be seen, the total area of $16\hbar \Theta$ this time is filled with only $8$ quantas, each of them having a different area state.}
\label{fig2}
\end{figure}

As mentioned before, we start by using the area law (Eq.\eqref{eq-2}) which is derived in the reference \cite{7}. The only difference is that we omit the constant $\pi$ as a matter of convenience. We assume each quanta of space can have a specific value of area which depends on its energy. So the area of each quanta of space can be written as
\begin{equation}\label{eq-6}
A_n = 2 \hbar \Theta(n + \frac{1}{2}).
\end{equation}
$n = 0$ corresponds to the ground state, which means there exist no energy in the space. In the result, space would be flat. Also, the space quantas area will all be the same when $n = 0$, having a value equal to $\hbar \Theta$ (Fig.\eqref{fig1}). Just by inserting some energy or mass into space, it's quantas would become excited, and instead of having quantas with the same area, we will see quants with different areas (Fig.\eqref{fig2}) depending on their energy. As we assumed the area of each
 quanta depends on its energy, so there must be a relation between energy $E_n$ and area $A_n$ of each space quanta, which we write in the following form,
\begin{equation}\label{eq-7}
E_n = \alpha A_n, \quad \text{with}\quad\alpha = \chi \Omega,
\end{equation}
where $\chi$ is an unknown dimensionless constant, and $\Omega$ is a constant to set the dimension. Since there exist no exact knowledge of the state of the space quantas, we should perform an ensemble average on all area states to find  the expectation value of area $\langle A \rangle$ of each quanta in the excited case. The expectation value for space quanta area reads
\begin{equation}\label{eq-8}
\langle A \rangle = \frac{\sum_{n=0}^{\infty}A_n e^{- \frac{E_n}{k_BT}}}{\sum_{n=0}^{\infty}e^{- \frac{E_n}{k_B T}}},
\end{equation}
where $k_B$ is the Boltzmann constant, and $T$ is the temperature.
By using Eq.\eqref{eq-6} and Eq.\eqref{eq-7}, we can write $\langle A \rangle = \frac{\sum_{n=0}^{\infty}A_n e^{-\beta A_n}}{\sum_{n=0}^{\infty}e^{-\beta A_n}}$, where $\beta = \frac{\alpha}{k_BT}$, and $\sum_{n=0}^{\infty}e^{-\beta A_n}$ is the partition function. By simplifying Eq.\eqref{eq-8}, the expectation value of area can be written as
\begin{equation}\label{eq-10}
\langle A \rangle = \hbar \Theta \Big( 1 + \frac{2}{e^{2\beta(\hbar \Theta)}-1} \Big).
\end{equation}
Also, If we assume no interaction between space quantas, and $N$ as the total number of quantas in the space, the total area $A_\sigma$ can be denoted by
\begin{equation}\label{eq-11}
A_\sigma = N\langle A \rangle.
\end{equation}
Now, using partition function, we can calculate the entropy of space. To do so, we calculate first the partition function of one space quanta in the form $Z_1 = \sum_{n=0}^{\infty}e^{-\beta A_n} = \frac{e^{-\beta \hbar \Theta}}{1 - e^{-2\beta \hbar \Theta}}$. As we assumed no interaction between quantas of space, the partition function of the whole space would be
$Z = (Z_1)^N = \Big( \frac{e^{-\beta \hbar \Theta}}{1 - e^{-2\beta \hbar \Theta}} \Big)^N$.
By using the system partition function, the Helmholtz free energy can be written as
\begin{equation}\label{eq-14}
F = -k_B T ln(Z) = k_B T N ln(e^{\beta \hbar \Theta} - e^{-\beta \hbar \Theta}).
\end{equation}
In the other hand, the relation $dF = - S dT + \sum_{i}X_i dY_i$ exists between thermodynamical variables of the system and Helmholtz free energy, where for constant $Y_i$'s relation between free energy and entropy would reduce to $S = -\frac{\partial F}{\partial T}$. With latter relation in hand, and Helmholtz free energy, entropy of space can be calculated as follows.
\begin{equation}\label{eq-17}
S = - k_B N \ln(e^{\beta \hbar \Theta} - e^{-\beta \hbar \Theta}) + \frac{\alpha N \hbar \Theta}{T} \Big( \frac{e^{2\beta \hbar \Theta} + 1}{e^{2\beta \hbar \Theta} - 1} \Big)
\end{equation}
By substituting Eq.\eqref{eq-10} in Eq.\eqref{eq-11}, and finally putting the result in Eq.\eqref{eq-17} we get
\begin{equation}\label{eq-18}
S = \frac{\alpha A_\sigma}{T} - A_\sigma \Big[ \frac{k_B}{\hbar \Theta}\Big( \frac{e^{2\beta \hbar \Theta} -1}{e^{2\beta \hbar \Theta} +1}\Big) \Big] \ln(e^{\beta \hbar \Theta} - e^{-\beta \hbar \Theta} ).
\end{equation}
To continue, we fix the dimension by multiplying RHS of Eq.\eqref{eq-18}, by $(\frac{k_B c^3}{ \hbar G})(\frac{\hbar G}{k_B c^3})$, and we get $S = (\frac{k_B c^3 A_\sigma}{\hbar G})(\frac{\hbar G}{k_B c^3})\Big(\frac{\alpha}{T} - \frac{k_B M}{\hbar \Theta}\Big)$, where $M$ is defined as a dimensionless function of temperature equal to
\begin{equation}\label{eq-21}
M = \Big( \frac{e^{2\beta \hbar \Theta} -1}{e^{2\beta \hbar \Theta} + 1}\Big) \ln(e^{\beta \hbar \Theta} - e^{-\beta \hbar \Theta} ).
\end{equation}
Since $(\frac{k_B c^3 A_\sigma}{\hbar G})$ has the dimension of entropy$[JK^{-1}]$, then $(\frac{\hbar G}{k_B c^3})(\frac{\alpha}{T} - \frac{k_B M}{\hbar \Theta})$, must be dimensionless. Also, from Eq\eqref{eq-6}, we see $\Theta$ must have the dimension of $[TM^{-1}]$, As a result $\alpha$ must have dimension of $[MT^{-2}]$.
In Short, using quantum mechanics and statistical mechanics we calculated the entropy of two dimensional quantized non-commuting space as below
\begin{equation}\label{eq-22}
S = (\frac{k_B c^3 A_\sigma}{\hbar G})\gamma(T)
\end{equation}
where $\gamma(T)$, is defined as
\begin{equation}\label{eq-23}
\gamma(T) = \Big(\frac{\hbar G}{k_B c^3}\Big)\Big(\frac{\alpha}{T}-\frac{k_B M}{\hbar\Theta}\Big).
\end{equation}
As the third law of thermodynamics states, when temperature goes to zero, the entropy must show the same behavior. So we are going to check this law for the Eq.\eqref{eq-22}. With a small calculation we can easily see, when $T \to 0$, in Eq.\eqref{eq-23}, $\gamma (T) \to 0$, which means the entropy tends to zero and Eq.\eqref{eq-23} obeys the third law of thermodynamics.

Now we proceed to calculate the constants $\Omega$ and $\Theta$ in terms of Planck units. Considering Eq.\eqref{eq-7}, since $\chi$ is dimensionless, $\Omega$ must have the dimension of $[MT^{-2}]$, we choose $\Omega = \frac{m_p}{t_p^2}$, where $m_p$ is the Planck mass and $t_p$ is the Planck time. Also, considering the dimension of $\Theta$, which is $[TM^{-1}]$, we choose $\Theta = \frac{t_p}{m_p}$, where $t_p$ and $m_p$, are again the Planck time and mass respectively and have the value $ t_p = \sqrt{\frac{\hbar G}{c^5}}$, and $m_p = \sqrt{\frac{\hbar c}{G}}$.
Using Eq.\eqref{eq-21}, we can write the constants $\Omega$ and  $\Theta$ as $
\Omega = \sqrt{\frac{c^{11}}{\hbar G^3}}$, $\alpha = \chi\sqrt{\frac{c^{11}}{\hbar G^3}}$, and $\Theta = \frac{G}{c^3}$.
In continue, we rewrite Eq.\eqref{eq-21} and Eq.\eqref{eq-23} in a simpler form, using the constants we found as
\begin{equation}\label{eq-26}
M = \Big( \frac{e^{\frac{2T_0}{T}} - 1}{e^{\frac{2T_0}{T}} + 1}\Big) \ln(e^{\frac{T_0}{T}} - e^{-\frac{T_0}{T}})
\end{equation}
\begin{equation}\label{eq-27}
\gamma(T) = (\frac{T_0}{T} - M)
\end{equation}
where $T_0$ is defined as $T_0 = \chi \sqrt{\frac{\hbar c^5}{Gk_B^2}} = \chi T_p$, where $T_p$ is the Planck temperature.
Now by using Eq.\eqref{eq-26}, we can write the inequality $e^{\frac{2T_0}{T}} - e^{-\frac{2T_0}{T}} > 0$. By solving this inequality we can finally obtain the condition $ T > 0$ for temperature. Besides, even in temperatures close to $T_0$ temperature $\frac{T}{T_0} \gg 1$, and we can expand Eq.\eqref{eq-26}, and by substituting the result in Eq.\eqref{eq-27}, the relation for entropy can be written as $S = (\frac{k_B c^3 A_\sigma}{\hbar G})\gamma(T)$, with
\begin{equation}\label{eq-32}
\gamma(T) = \frac{T_0}{T}\Big[ 1 - \ln(\frac{2T_0}{T}) \Big].
\end{equation}
Since Eq.\eqref{eq-32} must have a positive value, $(1 - \ln(\frac{2T_0}{T})) > 0$ always holds, and by a little algebra we can find $T \geq \frac{2T_0}{e}$.
Lastly, we have to mention that the calculated entropy is very similar to the entropy of the photon gas black-body $ s = (\frac{4 \pi ^2 k_B^4}{45 c^3 \hbar^3}) VT^3$\cite{8}.
Now we would use the Legendre transform between Helmholtz free energy and the system's internal energy to calculate the internal energy $U$.
The relation $F = U -TS$ exists between Helmholtz free energy $F$ and internal energy $U$. Now using Eq.\eqref{eq-14} and Eq.\eqref{eq-22}, The internal energy $U$ can be written in the form
\begin{equation}\label{eq-37}
U = (\frac{A_\sigma k_B c^3 }{\hbar G} T_0).
\end{equation}
We have also the first law of thermodynamics as the following.
\begin{equation}\label{eq-36}
dU = \partial W + \partial Q
\end{equation}

\subsection{Deriving EFE and Modified gravity equation}\label{part2}
In recent years, few scientists tried to derive the EFE using the first law of thermodynamics, where we can name the references \cite{9, 10} as most important efforts. In this section, we proceed to derive the EFE by using Eq.\eqref{eq-37}, and equations in references \cite{9, 10}. Since we did not assume any specific horizon, and the only condition we put on the space quantas is that quantas can be excited in presence of energy, and turn the flat space to a curved one, then our result is general and holds for any two-dimensional non-commuting space. So we are allowed to use the results derived in References \cite{9,10}, which have derived the EFE by considering local Rindler casual horizons.
To derive the EFE, we first differentiate the Eq.\eqref{eq-37}, and use the relations in references \cite{9,10}. In reference \cite{9}, $\partial Q$ and $\partial A_\sigma$ are defined as
\begin{equation}\label{eq-38}
\partial A_\sigma = - \eta \int \lambda R_{ab} k^a k^b d\lambda dA
\end{equation}
\begin{equation}\label{eq-39}
\partial Q = - \eta' \int \lambda T_{ab} k^a k^b d\lambda dA
\end{equation}
where $\eta$ and $\eta'$ are constants to set the dimension, $k^\mu$ is the tangent vector, and $T_{ab}$ and $R_{ab}$, are the energy-momentum and Ricci tensors respectively. By substituting Eq.\eqref{eq-37}, Eq.\eqref{eq-38} and Eq.\eqref{eq-39} in Eq.\eqref{eq-36}, we reach to the following relation.
\begin{align}\label{eq-40}
(\frac{T_0 k_B c^3}{\hbar G})(- \eta \int \lambda R_{ab} k^a k^b d\lambda dA) = &- \eta'\int \lambda T_{ab} k^a k^b d\lambda dA \nonumber \\&+ \partial W
\end{align}
To continue we use the relation between pressure $P$ and work $W$ which is in the form $
\partial W  = -P dV$. In the other hand, considering the relation between pressure $P$ and cosmological constant $\Lambda$ in the form $P = -\frac{\Lambda}{8\pi}$\cite{kubizvnak2012p}. We can conclude, that there must be a relation between work $W$, and cosmological constant $\Lambda$, which can be written in the following form
\begin{equation}\label{eq-43}
\partial W = \xi (\int \lambda (\frac{R}{2} - \Lambda)(g_{ab} k^a k^b d\lambda dA),
\end{equation}
where $\xi$ is a constant for dimension setting. Considering the reference \cite{9} (for all null vectors), the term $g_{ab}k^ak^b = 0$, so the work term in EFE is equivalent to zero, which might not be zero in modified gravity models.\\
Lastly, by substituting Eq.\eqref{eq-43} in Eq.\eqref{eq-40}, we find the next result.
\begin{equation}\label{eq-44}
(\frac{T_0 k_B c^3}{\hbar G}) \eta R_{ab} - \xi(\frac{1}{2} g_{ab} R - \Lambda g_{ab}) = \eta' T_{ab}
\end{equation}
To find a relation between the coefficients $\eta$, $\eta'$, and $\xi$, we expand Eq.\eqref{eq-44} in the weak filed limit. We expect to reach the Poisson equation, and we find a relation as follows.
\begin{equation}\label{eq-45}
\nabla ^2 \phi = \frac{\eta'}{2C} \Big( \frac{3\xi - 2C}{C - 2\xi} \Big) \rho c^4 + \frac{\xi}{C - 2\xi} \Lambda c^2,\quad C = (\frac{T_0 k_B c^3}{\hbar G}) \eta 
\end{equation}
By comparing the above result (Eq.\eqref{eq-45}) with $\nabla ^2 \phi = 4 \pi G \rho - \Lambda c^2$, we would have $\eta' = -\frac{8\pi G}{c^4}\Big( \frac{C(C - 2\xi)}{2C - 3\xi}\Big)$, and $C = \xi$. Now by substitution of $\eta'$ in Eq.\eqref{eq-44}, we find $R_{ab} - \frac{1}{2} g_{ab} R + \Lambda g_{ab} = \frac{8\pi G}{c^4} T_{ab}$, which is the EFE.

In recent years, different models of modified gravity have been suggested, which seem to be practical in solving new problems in cosmology, such as dark energy and dark matter. In continue, we try to investigate these modifications. Considering Eq.\eqref{eq-36}, \eqref{eq-38}, and\eqref{eq-39}, we can say that internal energy $U$, and thermal energy $Q$, has unique values, but this does not hold for work $W$. So we can use work $W$ in order to obtain different modification of gravity.\\
Let's rewrite the the work term in the form $\partial W = \zeta (\int \lambda (\frac{R}{2} - \Lambda)(g_{ab}k^ak^b d\lambda dA) + \int \lambda \chi_{ab} k^a k^b d\lambda dA)$, where $\zeta$ is a constant and $\chi_{ab} k^ak^b \neq 0$, is a parameter to describe equation of motion of different modifications of gravity like Lovelock gravity \cite{lovelock1971einstein}, massive gravity model \cite{de2011resummation}, or other models \cite{cardoso2018black}. So in result we can write our field equation as follows
\begin{equation}\label{eq-48}
R_{ab} - \frac{1}{2} g_{ab} R + \Lambda g_{ab} + \Phi \chi_{ab} = \frac{8\pi G}{c^4}T_{ab}
\end{equation}
where $\Phi$ in Eq.\eqref{eq-48}, is a constant.\\
We can conclude that each time physicists have tried to modify gravity to solve the new problems in Theoretical Physics, they have changed the work $W$ term in Eq.\eqref{eq-36}, and maybe we could find a unique and complete work term for gravity some day.

\subsection{Deriving Newton's law of gravity}\label{part3}
In this section, we are going to use the obtained relations to find an equation for gravity in quantum mechanical scale. For this sake, we first substitute Eq.\eqref{eq-10} in Eq.\eqref{eq-11}, which yields the following relation.
\begin{equation}\label{eq-49}
A_\sigma = N\hbar \Theta \frac{e^{2\beta \hbar \Theta} + 1}{e^{2\beta \hbar \Theta} - 1}
\end{equation}
We found the relation between total internal energy of the system $U$ with the 2 dimensional area $A_\sigma$ in Eq.\eqref{eq-37}. Now if we use what we found in Eq.\eqref{eq-49} in Eq.\eqref{eq-37} the result would be in the form of Eq.\eqref{eq-50}.
\begin{equation}\label{eq-50}
U = N k_B T_0 \Big( 1 + \frac{2}{e^{\frac{2T_0}{T}} - 1}\Big)
\end{equation}
In the limit $T \to 0$, which is equivalent to no energy existence on the space, the latter equation will become in the form of $U' = N k_B T_0$, where $k_BT_0$ denotes the energy of a single space quanta when there exists no energy on the space. So the energy of $N$ space quantas of an empty space would be equal to $Nk_BT_0$. Now we are able to subtract $U'$ from Eq.\eqref{eq-50} in order to find the energy of the objects on space separately.
\begin{equation}\label{eq-51}
	E = U - U' = \frac{2Nk_BT_0}{e^{\frac{2T_0}{T}} - 1}
\end{equation}

Now, if we rewrite Eq.\eqref{eq-49} in the form $N = \frac{A_\sigma}{\hbar \Theta} (\frac{e^{2\beta \hbar \Theta} -1}{e^{2\beta \hbar \Theta} + 1})$, then substitute in the Eq.\eqref{eq-51} the following relation would be obtained.
\begin{equation}\label{eq-52}
	k_BT=\frac{2k_BT_0}{\ln(\frac{bA_\sigma}{E} - 1)}\quad,\quad b = 2 \chi \sqrt{\frac{c^{11}}{\hbar G^3}}
\end{equation}
Also, Unruh has shown that a body with acceleration $a$ will experience a temperature $T$ from the background with the relation $k_B T = \frac{\hbar a}{2 \pi c}$ \cite{unruh1976notes}. Now, we can write the Eq.\eqref{eq-52} as follows,
\begin{equation}\label{eq-53}
k_BT = \frac{\hbar a}{2\pi c} \rightarrow a = \frac{4\pi c k_B T_0}{\hbar}\cdot\frac{1}{\ln (\frac{b A_\sigma}{E} - 1)}
\end{equation}
Eq.\eqref{eq-53} is an important relation which tells us an object with energy $E$ will excite space equal to area $A_\sigma - NA_0$, and in result of this excitation and deformation of the space, the object will have an acceleration equal to $a$. In continue, we will use Newton's second law ($F = ma$) and expansion of Eq.\eqref{eq-53} we will find the Newton's law of gravity.
If we expand the natural logarithm term in Eq.\eqref{eq-53} by using $\ln(-1+x) = - \sum_{k=1}^{\infty} \frac{(-1)^k(-2+x)^k}{k}$, which holds when $|-2+x| < 1$, we will have $\ln (\frac{bA_\sigma}{E} - 1) = \frac{bA_\sigma}{E} - 2$, then Eq.\eqref{eq-53} can be written in the following form.
\begin{equation}\label{eq-54}
	a \simeq \frac{4\pi c k_B T_0}{\hbar}\cdot \frac{1}{\frac{b A_\sigma}{E} - 2}
\end{equation}
If we consider the limit $\hbar \to 0$ in Eq.\eqref{eq-54}, we can use $E = mc^2$, then using Newton's second law of motion ($a=\frac{F}{m}$), we will end up with the following result.
\begin{equation}\label{eq-55}
	F \simeq \frac{1}{2}\frac{GmM}{r^2}
\end{equation}
This holds for the case $r \ll (\frac{M}{4\pi \chi}\sqrt{\frac{\hbar G^3}{c^7}})^{\frac{1}{2}}$. 
Paying attention to the fact that in the limit $\hbar \to 0$, we are out of quantum mechanical regime and we are allowed to use Newton's second law of motion. Eq.\eqref{eq-55} is very similar to what Newton suggested as the law of gravity, except in the $\frac{1}{2}$ coefficient. Our purpose in this paper is to understand the relation between gravity and quantum mechanics but not to present a rigorous formulation of the gravity. The coefficient $\frac{1}{2}$ in Eq.\eqref{eq-55} might disappear by changing the initial conditions on quantization of space. In addition, we do know that in the weak field approximation EFE reduces to Newton's law of gravity. In section \ref{part2} we derived EFE, and in section \ref{part3}, with another approach we derived Eq.\eqref{eq-55}, which is only different in a coefficient with Newton's law of gravity. It might be thought that Eq.\eqref{eq-55} is in contradiction with EFE, but we have to pay attention to the fact that in section \ref{part2} we used Newton's law of gravity to find the coefficient of Eq.\eqref{eq-40}, but if Eq.\eqref{eq-55} were used instead of Newton's law, the result were only different in a coefficient with EFE.

\section{Conclusion}

In this paper, without any tough and complicated calculations of high energy Physics, and just by using elementary equations of thermodynamics and non-commuting quantized space, we found a relation for entropy $S$ of space. Then by calculating the internal energy $U$, and using references \cite{9,10}, we derived the Einstein's field equation (EFE) and modified gravity equation. It is evident that all the efforts made in order to modify the Einstein's gravity have changed the work term $W$ in the
 Eq.\eqref{eq-36}. We also found Newton's law of gravity, using Holographic principle and Unruh effect. Just as Einstein claimed, we can argue that gravity is not a fundamental phenomenon. We assert that gravity depends on other fundamental variables such as the energy of space. Here we can ask the question, \emph{what is gravity?}
To answer this, we can say, gravity is nothing but the result of excitation of space quantas due to inserting energy on space. This excitation will cause the space quantas to have different areas and in the result, space will become curved.

\section{Aknowledgments}
The authors want to express their gratitude to Mostafa Mahdavi, Hamidreza Salahi, and Ali Nezhadsafavi for their collaborative talks. Also, we want to thank anyone who helped in the process of preparing this paper.


\begin{thebibliography}{10}

\bibitem{blum2015brief}
A.~Blum, ``A brief history of string theory: From dual models to m-theory,''
  2015.

\bibitem{damour1994string}
T.~Damour and A.~M. Polyakov, ``String theory and gravity,'' {\em General
  Relativity and Gravitation}, vol.~26, no.~12, pp.~1171--1176, 1994.

\bibitem{seiberg1999string}
N.~Seiberg and E.~Witten, ``String theory and noncommutative geometry,'' {\em
  Journal of High Energy Physics}, vol.~1999, no.~09, p.~032, 1999.

\bibitem{rovelli2008loop}
C.~Rovelli, ``Loop quantum gravity,'' {\em Living reviews in relativity},
  vol.~11, no.~1, p.~5, 2008.

\bibitem{carlip2001quantum}
S.~Carlip, ``Quantum gravity: a progress report,'' {\em Reports on Progress in
  Physics}, vol.~64, no.~8, p.~885, 2001.

\bibitem{1}
J.~D. Bekenstein, ``Black holes and entropy,'' {\em Physical Review D}, vol.~7,
  no.~8, p.~2333, 1973.

\bibitem{2}
S.~W. Hawking, ``Black hole explosions?,'' {\em Nature}, vol.~248, no.~5443,
  p.~30, 1974.

\bibitem{9}
T.~Jacobson, ``Thermodynamics of spacetime: the einstein equation of state,''
  {\em Physical Review Letters}, vol.~75, no.~7, p.~1260, 1995.

\bibitem{11}
E.~Verlinde, ``On the origin of gravity and the laws of newton,'' {\em Journal
  of High Energy Physics}, vol.~2011, no.~4, p.~29, 2011.

\bibitem{bousso2002holographic}
R.~Bousso, ``The holographic principle,'' {\em Reviews of Modern Physics},
  vol.~74, no.~3, p.~825, 2002.

\bibitem{gouba2016comparative}
L.~Gouba, ``A comparative review of four formulations of noncommutative quantum
  mechanics,'' {\em International Journal of Modern Physics A}, vol.~31,
  no.~19, p.~1630025, 2016.

\bibitem{snyder1947quantized}
H.~S. Snyder, ``Quantized space-time,'' {\em Physical Review}, vol.~71, no.~1,
  p.~38, 1947.

\bibitem{connes1985non}
A.~Connes, ``Non-commutative differential geometry,'' {\em Publications
  Math{\'e}matiques de l'Institut des Hautes {\'E}tudes Scientifiques},
  vol.~62, no.~1, pp.~41--144, 1985.

\bibitem{akofor2008quantum}
E.~Akofor, A.~Balachandran, and A.~Joseph, ``Quantum fields on the
  groenewold--moyal plane,'' {\em International Journal of Modern Physics A},
  vol.~23, no.~11, pp.~1637--1677, 2008.

\bibitem{7}
J.~M. Romero, J.~Santiago, and J.~D. Vergara, ``Note about the quantum of area
  in a noncommutative space,'' {\em Physical Review D}, vol.~68, no.~6,
  p.~067503, 2003.

\bibitem{bekenstein1974quantum}
J.~D. Bekenstein, ``The quantum mass spectrum of the kerr black hole,'' {\em
  Lettere al Nuovo Cimento (1971-1985)}, vol.~11, no.~9, pp.~467--470, 1974.

\bibitem{alekseev2003area}
A.~Alekseev, A.~Polychronakos, and M.~Smedb{\"a}ck, ``On area and entropy of a
  black hole,'' {\em Physics Letters B}, vol.~574, no.~3-4, pp.~296--300, 2003.

\bibitem{8}
H.~S. Leff, ``Teaching the photon gas in introductory physics,'' {\em American
  Journal of Physics}, vol.~70, no.~8, pp.~792--797, 2002.

\bibitem{10}
C.~Eling, ``C. eling, r. guedens, and t. jacobson, phys. rev. lett. 96, 121301
  (2006).,'' {\em Phys. Rev. Lett.}, vol.~96, p.~121301, 2006.

\bibitem{kubizvnak2012p}
D.~Kubiz{\v{n}}{\'a}k and R.~B. Mann, ``P- v criticality of charged ads black
  holes,'' {\em Journal of High Energy Physics}, vol.~2012, no.~7, p.~33, 2012.

\bibitem{lovelock1971einstein}
D.~Lovelock, ``The einstein tensor and its generalizations,'' {\em Journal of
  Mathematical Physics}, vol.~12, no.~3, pp.~498--501, 1971.

\bibitem{de2011resummation}
C.~De~Rham, G.~Gabadadze, and A.~J. Tolley, ``Resummation of massive gravity,''
  {\em Physical Review Letters}, vol.~106, no.~23, p.~231101, 2011.

\bibitem{cardoso2018black}
V.~Cardoso, M.~Kimura, A.~Maselli, and L.~Senatore, ``Black holes in an
  effective field theory extension of general relativity,'' {\em Physical
  review letters}, vol.~121, no.~25, p.~251105, 2018.

\bibitem{unruh1976notes}
W.~G. Unruh, ``Notes on black-hole evaporation,'' {\em Physical Review D},
  vol.~14, no.~4, p.~870, 1976.

\end{thebibliography}
\end{document}